\documentclass[aps,prl,twocolumn,groupedaddress]{revtex4}
\usepackage{graphicx}
\usepackage{amsmath}
\usepackage{amssymb}

\begin{document}

\title{Langevin dynamics neglecting detailed balance condition}
\author{Masayuki Ohzeki}{
\affiliation{Department of Systems Science, Kyoto University, Yoshida-Honmachi, Sakyo-ku, Kyoto 606-8501, Japan}
\author{Akihisa Ichiki}
\affiliation{Green Mobility Collaborative Research Center, Nagoya University, Furo-cho, Chikusa-ku, Nagoya 464-8603, Japan}
\date{\today }

\begin{abstract}
An improved method for driving a system into a desired distribution, for example, the Gibbs-Boltzmann distribution, is proposed, which makes use of an artificial relaxation process.
The standard techniques for achieving the Gibbs-Boltzmann distribution involve numerical simulations under the detailed balance condition.
In contrast, in the present study we formulate the Langevin dynamics, for which the corresponding Fokker-Planck operator includes an asymmetric component violating the detailed balance condition. 
This leads to shifts in the eigenvalues and results in the acceleration of the relaxation toward the steady state.
The numerical implementation demonstrates faster convergence and shorter correlation time, and the technique of  biased event sampling, Nemoto-Sasa theory, further highlights the efficacy of our method.
\end{abstract}

\maketitle

{\it Introduction.---}
In order to estimate the physical quantities in the equilibrium state of a many-body system, we often perform a stochastic numerical simulation governed by the master or Fokker-Planck equations.
In numerical simulation in stochastic dynamics, however, an unavoidable obstacle (in the form of critical slowing down in the relaxation to the equilibrium state), is involved in the frustrated dynamics appearing in structural glassy systems, protein-folding simulations, and various critical phenomena.
In order to overcome this difficulty, researchers have proposed many alternatives beyond the standard approaches \cite{Swendsen1987,Hukushima1996,Neal2001,Ohzeki2010a}.
The majority of these methods, for instance, the Markov-chain Monte Carlo method (MCMC), adhere to the detailed balance condition (DBC), which is a simple solution satisfying the balance condition (BC) to assure relaxation to the equilibrium state.
However, it is not necessary that the DBC should be satisfied in order to generate the desired distribution.
Several ingenious techniques realize relaxation to the target steady state without satisfying the DBC  \cite{Suwa2010,Turitsyn2011,Fernandes2011} in the MCMC. 
In particular, Suwa and Todo have proposed a trick to severely reduce the rejection rate in an extension of stochastic dynamics that allows violation of the DBC.
This approach demonstrates faster convergence to the target steady state and reduces the correlation time, which is closely related to the number of samplings required to efficiently compute the expectation values.
However in these studies, the violation of the DBC is limited to the local part of the stochastic dynamics.
They have failed to change the global modification to eliminate the critical slowing down, which hampers the investigation of the many-body system.  

The above-listed studies are performed at the level of the master equation, which is a coarse-grained picture of the dynamics.
Direct observation of the stochastic behavior is performed by the Langevin dynamics, which is used for investigating the structural glassy system and protein folding.
Nowadays, the Langevin dynamics is available for optimization in the context of machine learning for big data due to its ease of implementation \cite{Welling2011}.
We naturally expect that modified microscopic dynamics that are free from the DBC exist; however, discussion on microscopic dynamics that violate the DBC is not widespread in the literature.
For example, one prominent question has not yet been addressed, i.e., the identification of forces that can accelerate the relaxation to the steady state without any change in the final distribution.
In the present study, we propose modified microscopic dynamics governed by the Langevin equation, with high-speed relaxation to the steady state with the desired distribution.
The proposed Langevin dynamics have a connection to the Fokker-Planck equation, neglecting the DBC.
Analysis of the eigenvalues of the corresponding Fokker-Planck operator theoretically assures the acceleration of the relaxation to the desired distribution, according to the same line of argument as that given in our preceding study \cite{Ichiki2013}.
We numerically demonstrate that our method actually accelerates the relaxation to the steady state while eliminating the critical slowing down at the critical temperature and confirm that  a reduction in the auto-correlation time is achieved.

{\it Duplicated system without DBC.---}
In order to simply formulate the Langevin dynamics, neglecting the detailed balance condition, we use a duplicated system as in the skewed detailed balance condition \cite{Turitsyn2011}.
The overdamped Langevin equation for the duplicated system with an identical isothermal heat bath with a temperature $T$ is defined as
\begin{equation}
d{\bf x}_i = {\bf A}_i dt + \sqrt{2T} d{\bf W}_i,
\end{equation}
where ${\bf x}_i$ represents the $N$-dimensional microscopic degrees of freedom of the $i$th system and $d{\bf W}_i$ is the $N$-dimensional Wiener process.
The quantity ${\bf A}_i$ represents a $N$-dimensional force.
Here, we do not adhere to the standard equilibrium form as ${\bf A}_i = - {\rm grad}_i U({\bf x}_i)$, where $U({\bf x}_i)$ is the identical potential energy in the duplicated system.

The corresponding Fokker-Planck equation for a duplicated system with $2 \times N$ particles, is given as
\begin{equation}
\frac{\partial}{\partial t}P({\bf x}_1,{\bf x}_2,t) = - \sum_{i=1,2}{\rm div}_{i} {\bf J}_i({\bf x}_1,{\bf x}_2),\label{FK}
\end{equation}
where $P({\bf x}_1,{\bf x}_2,t)$ is the time-dependent distribution, and the probabilistic flow, ${\bf J}_i({\bf x}_1,{\bf x}_2)$, is defined as
\begin{equation}
{\bf J}_i({\bf x}_1,{\bf x}_2) = \left\{ {\bf A}_i - T {\rm grad}_{i} \right\}P({\bf x}_1,{\bf x}_2,t),\label{flow}
\end{equation}
where the gradient and divergence appearing below with the subscript $i$ is taken for each system.

In order to generate the desired distribution after relaxation, we impose the distribution in the steady state as $P_{\rm ss}({\bf x}_1,{\bf x}_2,t) \propto \exp(- \sum_{i=1,2}U({\bf x}_i)/T)$.
Then the Fokker-Planck equation demands the following condition
\begin{equation}
0 = - \sum_{i=1,2}{\rm div}_{i} {\bf J}_i({\bf x}_1,{\bf x}_2).\label{FK2}
\end{equation}
A trivial solution with ${\bf J}_i=0$ yields the equilibrium system with ${\bf A}_i = - {\rm grad}_i U({\bf x}_i)$.
In the present study, we seek a nontrivial solution, different from the ordinary force, with an additional force as
\begin{equation}
{\bf A}_i = - {\rm grad}_{i} U({\bf x}_i) + \gamma {\bf f}_i,
\end{equation}
where $\gamma$ is an arbitrary parameter controlling the degree of violation of the DBC, as explained below.
By substituting the above expression into the condition (\ref{FK2}), we find that the additional force must satisfy $\sum_{i=1,2} {\rm div}_{i} \left( {\bf f}_i P_{\rm ss}({\bf x}_1,{\bf x}_2,t)\right) =0$.
A trivial solution on this condition is ${\bf f}_i = {\bf 1}\exp\left(U({\bf x}_i)/T\right)$ where {\bf 1} is a vector, with all the elements being unity.
The trivial solution is, however, problematic to implement since the force is unidirectional and includes an exponential term that is dependent on the temperature.
One may find a kind of the rotational force satisfying the divergence-free condition as
\begin{equation}
[{\bf f}_i]_k = \frac{\partial U({\bf x}_i)}{\partial [{\bf x}_i]_{k-1}} - \frac{\partial U({\bf x}_i)}{\partial [{\bf x}_i]_{k+1}},\label{rotf}
\end{equation}
where $[\cdot]_k$ denotes the $k$th element of the vector.
Unfortunately, the rotational force is not user-friendly for implementation of MCMC due to existence of the arbitrariness as $([{\bf x}_i]_k \to [{\bf x}_i]_{\pi(k)})$, where $\pi(\cdot)$ denotes a permutation of the elements.
This is the reason why we implement the duplicated system to introduce a nontrivial but simple rotational force as detailed below.
We introduce the following nontrivial solution
\begin{eqnarray}
{\bf f}_1 &=& {\rm grad}_2 U({\bf x}_2) \label{rep1},\\
{\bf f}_2 &=& - {\rm grad}_1 U({\bf x}_1) \label{rep2}.
\end{eqnarray}
These forces lead to a type of the mixture of the duplicated system, with the steady state unchanged.
Below we confirm that the additional force actually violates the DBC and accelerates the relaxation.

Let us calculate the transition probability for each system $(i=1,2)$ during an infinitesimal time interval, $[t,t+dt]$.
We obtain
\begin{eqnarray}\nonumber
&&L^{\gamma}_i({\bf x}_i(t+dt)|{\bf x}_i(t)) \\
&&\propto \exp\left\{-\frac{1}{4T}(\dot{{\bf x}}_i - {\bf A}_i)^2dt -\frac{1}{2}{\rm div} {\bf A}_i dt\right\},\label{transition}
\end{eqnarray}
where ${\bf x}_i(t)$ is the location at time $t$ for each system.
We omit the time dependence in the quantities on the right-hand side for simplifying the expressions.
We use the midpoint prescription, $\bar{{\bf x}}_i = ({\bf x}_i(t+dt)+{\bf x}_i(t))/2$; we take the partial derivative with respect to the location.
The ratio of the transition probability between the forward and backward processes confirms violation of the DBC due to the existence of the probabilistic flow for each system.
In this sense, $\gamma$ controls the degree of violation of the DBC.

Let us rewrite Eq. (\ref{FK}) using the ordinary Fokker-Planck operator for each system \cite{Risken1996}, where
\begin{eqnarray}
\left[{\bf a}_i \right]_k &=& \frac{1}{2\sqrt{T}}\frac{\partial U({\bf x}_i)}{\partial [{\bf x}_{i}]_k} + \sqrt{T}\frac{\partial}{\partial [{\bf x}_i]_k},
\\
\left[ {\bf a}_i^{\dagger} \right]_k &=& \frac{1}{2\sqrt{T}}\frac{\partial U({\bf x}_i)}{\partial [{\bf x}_i]_k} - \sqrt{T}\frac{\partial}{\partial [{\bf x}_i]_k}.
\end{eqnarray}
These operators satisfy $\left[ [{\bf a}_i]_k,[{\bf a}_j^{\dagger}]_l \right] = - \delta_{ij}\partial^2 U({\bf x}_i)/\partial[{\bf x}_i]_k\partial[{\bf x}_i]_l$, where the brackets without a subscript denote commutation.
We can then rewrite Eq. (\ref{FK}) using the above operators for each system as
\begin{eqnarray}\nonumber
& &\frac{\partial}{\partial t}\bar{P}({\bf x}_1,{\bf x}_2,t) \\ \nonumber
& &  = - \left\{ \sum_{i=1,2}{{\bf a}^{\dagger}}^{\rm T}_i {\bf a}_i - \gamma \left( {{\bf a}^{\dagger}_2}^{\rm T}{\bf a}_1 - {{\bf a}_1^{\dagger}}^{\rm T}{\bf a}_2 \right)\right\}\bar{P}({\bf x}_1,{\bf x}_2,t),\\
\end{eqnarray}
where $\bar{P}({\bf x}_1,{\bf x}_2,t) = \exp(\sum_{i=1,2}U({\bf x}_i)/2T)P({\bf x}_1,{\bf x}_2,t)$.
The steady state is given by the eigenfunction, $\varphi_0({\bf x}_1)\varphi_0({\bf x}_2)$, for $-{{\bf a}_i^{\dagger}}^{\rm T}{\bf a}_i~(i=1,2)$ whose eigenvalue vanishes because $\varphi_0({\bf x}_i)$ is the eigenfunction with zero eigenvalue for ${\bf a}_i~(i=1,2)$.
The term added to the ordinary Fokker-Planck operators comes from the additional force $\gamma {\bf f}_i$, which generates nonzero current of each system in the steady state.  

As discussed in Ref. \cite{Ichiki2013}, while the symmetric component, $\propto \sum_{i}{{\bf a}_i^{\dagger}}^{\rm T}{\bf a}_i$, is fixed, the introduction of the anti-symmetric component, $\propto {{\bf a}^{\dagger}_2}^{\rm T}{\bf a}_1 -{{\bf a}^{\dagger}_1}^{\rm T}{\bf a}_2$, can accelerate the relaxation, since the gap between the first and second eigenvalues increases.
Although we demonstrated this fact in finite dimensions in the previous study, it holds even in infinite dimensions, since the proof is straightforwardly extendable to the infinite dimensional case.

{\it As a biased sampling.---}
The introduction of the additional force can be interpreted as a type of the biased sampling, as proposed in Nemoto-Sasa theory and its generalization \cite{Nemoto2011,Sughiyama2013}.
The ratio of the modified and unmodified path probabilities is
\begin{eqnarray}
& &\ln\left\{\prod_{i=1,2}\frac{L^{\gamma}_i({\bf x}_i(t+dt)|{\bf x}_i(t))}{L^0_i({\bf x}_i(t+dt)|{\bf x}_i(t))}\right\}=\psi_\gamma dt,
\end{eqnarray}
where
\begin{eqnarray}\nonumber
& &\psi_\gamma =   \frac{\gamma}{2T}{\bf f}_i^{\rm T}\left[\dot{\bf x}_i+{\rm grad}_i U-\gamma {\bf f}_i\right] -\frac{\gamma^2}{4T} {\bf f}_i^2-\frac{\gamma}{2}{\rm div}{\bf f}_i.\\
\end{eqnarray}
Using Eq. (\ref{transition}), we can evaluate the expectation of $\psi_\gamma$ under dynamics with the DBC nonpositive as $\langle \psi_{\gamma} \rangle_{\gamma=0} \le 0$ while $\langle \psi_{\gamma} \rangle_{\gamma} \ge 0$.
Supposing the transition ${\bf x}_i\to{\bf x}'_i$ is typical under the dynamics with DBC, this fact implies that the probability for the same transition is exponentially reduced under the modified dynamics. 
If the typical transition, for instance trap in local minimum of potential energy, is a bottleneck in the convergence toward the equilibrium state, the violation of the DBC significantly accelerates the relaxation through the modified path when the dynamics is guaranteed to converge to the same distribution. 
This is the physical interpretation of the acceleration of the relaxation to the steady state.

{\it Demonstrations.---}
We first test our method in the double-valley potential for the one-dimensional system defined as
\begin{equation}
U(x) = -\frac{1}{2}x^2 + \frac{1}{4}x^4.
\end{equation}
The initial condition is set to be in one of the valleys at $x_0 = 1$.
The particle must go beyond the potential swelled at $x=0$.
The time evolution of the Langevin equation is evaluated using the ordinary method known as the Heun scheme \cite{Kloeden2011}.
We set the infinitesimal time as $d t = 0.0001$ and test two cases with $\gamma=0$ and $\gamma=10.0$.
We set the temperature as $T=1$.
The figure \ref{fig1} shows the results averaged over independent runs $N_{\rm sam}=1000$, while taking the time average during $\Delta t = 0.1$.
As mentioned in the previous section, the averaged orbits over $N_{\rm sam} = 1000$ differ depending on the value of $\gamma$.
This fact ensures that the change in the typical behavior indeed occurs.
We confirm the faster convergence to the desired distribution in terms of the correct estimation of the expectation of the location.
In addition, we compute the (integrated) auto-correlation time $\tau_{\rm int}$ defined as $\tau_{\rm int} = \sum_{t'=1}^{\infty}\left(\langle x(t) x({t+t'})\rangle - \langle x \rangle^2\right)/\left(\langle x^2 \rangle - \langle x \rangle^2\right)$, where $x(t)$ is the location of the particle at time $t$ and the bracket denotes the ensemble average.
We compute it by omitting the first relaxation and taking the average over several $t$ to eliminate the dependence on $t$.

We confirm reduction of $\tau_{\rm int}$ for each $\gamma$ as $1.27 (\gamma=0.0)$, $0.67~(\gamma=1.0)$, $0.31~(\gamma=2.0)$,  $0.13~(\gamma=5.0)$, and $0.06~(\gamma=10.0)$.
All of the results shown above ensure that our method actually accelerates the relaxation toward the steady state and further makes the correlation time shorter by induction of the additional force.
\begin{figure}[tb]
\begin{center}
\includegraphics[width=0.45\textwidth]{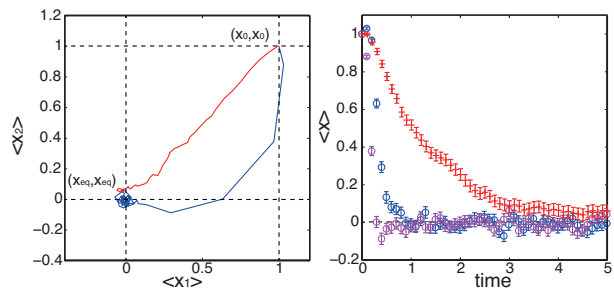}
\end{center}
\caption{ (Color online)
(Left) Orbits of the duplicated system until $t=5$. 
The horizontal and vertical axes denote the mean of the location.
The red curve denotes the case of $\gamma=0$ and the blue one represents that of $\gamma=10.0$. 
(Right) Time evolution of the mean of the location.
The horizontal axis represents time, and the vertical axis denotes the mean of the location.
From top to bottom, we plot the cases of $\gamma=0$ (red crosses) and $\gamma = 10.0$ (blue and purple circles).
}
\label{fig1}
\end{figure}

Next, we demonstrate the significant acceleration of the relaxation to the steady state by showing the removal of the critical slowing down of the XY model.
The potential energy of the XY model is defined as
\begin{equation}
U({\bf x}) = - \sum_{j=1}^N\sum_{k \in \partial j}\cos\left(x_j -x_k\right),
\end{equation}
where the summation is taken over the adjacent pairs to the spin, $j$, on the square lattice with a periodic boundary condition.
Note that ${\bf x}$ here denotes the spin directions on the lattice such that ${\bf x} \in [0,2\pi)^{N}$ on each site.
The XY model does not exhibit any spontaneous symmetry breaking even in two dimensions, but the so-called Kosterlitz-Thouless (KT) transition occurs at $T_c = 0.89213(10)$ \cite{Olsson1995}.
In the low-temperature region (KT phase), the magnetization relaxation exhibits critical power-law behaviour \cite{Nishimori2011} and oscillates around $m=0$ owing to the finite-size effect.
This implies that the magnetization trivially vanishes in the equilibrium state, but the dynamical behaviour during relaxation is not trivial.
The number of degrees of freedom is set to $N=10\times 10$ and that of the independent runs is $N_{\rm sam} = 1000$.
We set the temperature as $T=0.5$ below $T_c$.

The initial condition that all spins are in the `up'-direction, i.e., ${\bf x}_i = \pi/2$ for $i=1,2$, is imposed.
We observe the relaxation of the magnetization $m = \sum_{j=1}^N\sin x_j/N$ and internal energy while taking the time average during $\Delta t = 0.1$.
The obtained data of $m$ oscillates because of the critical behavior in the finite-size system.
We then take the mean of the independent runs.
The variance of the observed data is attributed to this oscillating behavior.
We observe critical slowing down in the KT phase in the relaxation for the case of $\gamma=0$, as shown in Fig. \ref{fig2}.
On the other hand, the case of $\gamma=10$ does not show critical slowing down in the KT phase.
This fact implies that the additional force significantly accelerates the relaxation to the steady state.
This is a remarkable point of our method.
The other method without DBC as the Suwa-Todo method and skewed DBC \cite{Suwa2010,Turitsyn2011,Fernandes2011} is based on the ``local" tuning of the transition rule.
However our method ``globally" changes the driving force in the system.
This is the reason why the critical slowing down is eliminated.
\begin{figure}[tb]
\begin{center}
\includegraphics[width=0.45\textwidth]{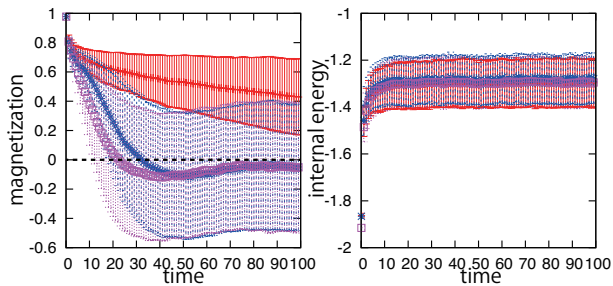}
\end{center}
\caption{ (Color online) relaxation in magnetization (left panel) and internal energy (right panel) in KT phase.
The horizontal axis represents time, and the vertical axis represents magnetization and internal energy.
The cases of $\gamma = 0$ (red crosses) and $\gamma = 10.0$ (blue tilted crosses and purple squares) are plotted.
}
\label{fig2}
\end{figure}
We also observe the internal energy (Fig. \ref{fig2}).
We confirm that consistent values are obtained independent of the gamma value.
This observation verifies that the additional force does not change the steady state.

We must point out that some numerical techniques may be necessary in order to implement our method.
A small $\gamma$ value is required depending on the scheme to implement the stochastic dynamics.
In the present study, we employ the Heun scheme, but the Euler-Maruyama scheme is too poor in precision to be utilized for a large value of the force.
In addition, some readers might think that our method resembles the replica-exchange Monte Carlo method \cite{Hukushima1996}.
In this method, we prepare several replicas of the system with small-different temperatures to support the stochastic jump from the valley of the potential energy.
However our method does not necessarily demand replicas if we use the nontrivial solution (\ref{rotf}) and utilizes only two at most. 
In addition, one may implement our method in conjunction with the replica-exchange Monte Carlo method.
They are not competing.

{\it Conclusions.---}
In the present study, we propose a simple method for accelerating the relaxation to the desired distribution in the Langevin dynamics by introducing an additional force to violate the DBC.
We confirm that our method can actually accelerate the relaxation to the steady state for a double-valley system and the XY model on a square lattice.
In particular, the latter model involves critical slowing down in the low-temperature region, namely, the KT phase.
Our method demonstrates remarkable performance, escaping the potential valley and avoiding critical slowing down, while the steady state remains unchanged.

In order to implement the nontrivial solution of the additional force, we introduce the duplicated system.
The number of ``replicas" is not limited.
If we find a nontrivial solution for the replicated system with different heat baths, one can develop an excellent method, inspired by the replica exchange Monte Carlo simulation, in the Langevin dynamics in a relatively simple way.
Our method is based on mathematical assurance, in the form of a shift in the eigenvalue of the corresponding Fokker-Planck operator and the biased sampling.
Recent development reveals our additional force is a kind of optimal solution in the biased sampling \cite{Ichiki2015}. 
We hope that various implementations of such designed algorithms will become widely used in nonequilibrium statistical mechanics studies in the future. 
\begin{acknowledgments}
One of the authors (M.O.) is grateful for fruitful discussions with H. Suwa, K. Hukushima, M. Kikuchi, and H. Touchette.
The present study was inspired by a lecture given by J. Teramae. 
This work was supported by MEXT in Japan: KAKENHI No.24740263 and No.15H03699, and the Kayamori Foundation of Informational Science Advancement.
\end{acknowledgments}

\bibliography{paper_ver10}
\end{document}